\documentclass[preprint]{aastex}

\begin{document}

\title{The distances to open clusters
from main-sequence fitting. I. New models and a comparison
to the properties of the Hyades eclipsing binary vB~22}

\author{Marc H. Pinsonneault and Donald M. Terndrup} 
\affil{Ohio State University, Department of
Astronomy, Columbus, OH 43210} 
\email{pinsono,terndrup@astronomy.ohio-state.edu}

\author{Robert B. Hanson} 
\affil{University of California Observatories/Lick 
Observatory, Santa Cruz, CA 95064}
\email{hanson@ucolick.org}

\and

\author{John R. Stauffer} 
\affil{Infrared
Processing and Analysis Center, California Institute of
Technology, Mail Code 100-22, 770 South Wilson Avenue,
Pasadena, CA 91125}
\email{stauffer@ipac.caltech.edu}

\slugcomment{}

\shorttitle{Hyades isochrone}
\shortauthors{Pinsonneault et al.}

\begin{abstract} 
In the first of a new series
of papers on open cluster distances, we use updated 
stellar evolution models to construct an isochrone appropriate
for the Hyades, and compare it with 
the Hyades eclipsing binary system 
vB~22.  We find that the absolute and relative luminosities of
the two stars are in good agreement with the
model, but the radii do
not match the values inferred from eclipse data.
We present evidence that there is a consistency problem
with the flux ratios and the inferred radii, and discuss
possible theoretical effects that could be responsible for
the mismatch in the radii.  We derive a helium abundance
for the Hyades of $Y = 0.271 \pm 0.006$, which is
equal within the errors to the Sun's initial helium
abundance even though the Hyades is considerably more
metal-rich.
\end{abstract}


\section{Introduction}

The distances to Milky Way star clusters as derived
from main-sequence fitting play a critical role in
unraveling the history of the Galaxy and, via
luminosity calibration of pulsating variables,
in finding distances throughout the Local Group.   
Parallaxes from the Hipparcos
satellite have provided precise measurements of the
distances to the nearest open clusters, particularly
the Hyades \citep{per98}, allowing stringent tests of
the predictions of stellar evolutionary models
\citep[e.g.,][]{lebreton01}.  

To go beyond the handful of clusters with trigonometric 
parallaxes \citep[e.g.,][]{vl99,rob99,makarov02} requires
isochrones that are physically accurate and 
well calibrated over a wide range of temperature --
ideally all along the main sequence.  As is well 
known this is a nontrivial
exercise, since the luminosity of the main-sequence is
a sensitive function of helium abundance and metallicity.
Radii from stellar models depend on the treatment of
convection, for which only simple phenomenological theories
are available.  Photometric colors and bolometric corrections are
often poorly determined, especially for stars much hotter
or cooler than the Sun, and are highly dependent on the
details of model atmospheres employed in the computation
of the isochrones.  Rapidly rotating stars typically
have large spots and chromospheric emission, not modelled in the
computation of the isochrones, which could
affect their colors especially in the blue and ultraviolet
\citep{vla87,sta03}, As a result of some or
all these effects, it is typically the case
that even the best isochrones have do not match the detailed shape
of the main sequence as determined from photometry 
\citep{ter00,cas01,lebreton01}.

Despite the complexity of the problem, it is now possible
to determine more accurate absolute and relative
distances to open clusters.  Helioseismology 
has given us reliable measures of many parameters which 
directly or indirectly affect the radius, including the 
helium abundance, the amount of helium diffusion, 
and opacities in the convective zone \citep[e.g.,][]{bpb01}.
Hipparcos parallaxes in the Hyades \citep{per98,debruijne01}
and of nearby field stars \citep{jimenez03,per03} can potentially
provide the means of empirically correcting the 
isochrone color-temperature relation.

In addition, eclipsing binary stars provide a powerful test of the
theory of stellar structure and evolution,
particularly the mass/luminosity relation.  This is
especially true of systems in star clusters, where there
are additional constraints on the age and abundances of
the stars.  As summarized by \citet{lebreton01},
the Hyades cluster has five binaries where
the components have measured masses.  Of these systems
only vB~22 has masses with a small enough uncertainty to
place powerful constraints on the theoretical models; we
will therefore focus on vB~22 (= 818~Tau, HD~27130).  In 
this system,
the relative magnitudes of vB~22A and vB~22B have been
measured in several colors and the absolute radii have
been inferred. 

We will use vB~22 as a test of both the absolute
luminosities and effective temperatures of our models.
We contend that the agreement in luminosity that we obtain
justifies the construction of an empirical isochrone where
the colors as a function of $M_V$ are adjusted to reproduce
the morphology of the Hyades color-magnitude diagram.
In addition, we will show that the absolute magnitudes of
the two components in the $B$, $V$, and $I$ bands provide support
for the relative model luminosities and effective temperatures,
even though the direct radii inferred from the binaries
are not in agreement with the models.  This step justifies
holding the model effective temperatures fixed and varying
the color calibrations when constructing the empirical
isochrone, the details of which are discussed
in the second paper in this series. Finally, we examine the question
of the Hyades helium abundance and the ratio $\Delta Y /
\Delta Z $ appropriate for chemical evolution of solar-neighborhood
stars.

The vB~22 system has been studied extensively since \citet{mcclure82}
first used it to determine the distance to the Hyades and to
constrain the mass-luminosity relationship.  The most recent
papers \citep{lastennet99,lebreton01,tr02} have yielded
somewhat discordant results. \citet{lebreton01} claim evidence for
a low helium abundance, while \citet{lastennet99}
and \citet{tr02} find that the luminosities of
the models are consistent with the data if the Hipparcos
distance is adopted.  All authors note the apparent
contradiction between the radii of theoretical models and
those obtained from the eclipse data.  In light of these
results we believe that a careful analysis of the data
and the theoretical models is warranted, with particular
attention to the errors involved. 

\section{An Isochrone for the Hyades}

We begin with a new set of theoretical models described in
detail by \citet{spt00}.  The essential aspects of these 
models are repeated here for the reader's convenience.

We used the Yale Rotating Evolution Code (YREC) to
construct evolutionary tracks over the mass
range $0.25 \leq (M / M_\odot) \leq 2.25$.  YREC is a Henyey
code which solves the equations of stellar structure in
one dimension \citep{guenther92} and which follows
rotational evolution by treating the star as a
set of nested, rotationally deformed shells. For this
application, however, we used the code in its
non-rotating mode; non-rotating stars of the age and masses
considered here are structurally identical to those
which are rotating \citep[e.g.,][]{spt00} and solar-like Hyades
stars are slow rotators \citep{radick87,psc03}.  The chemical
composition of each shell is updated separately using
the nuclear reaction rates of \citet{gb98}.  Composition
changes due to microscopic diffusion can be calculated.  The initial
chemical mixture is the solar mixture of \citet{gn93}, and
for the Sun the models have a surface metallicity of $Z =
0.0176$ at the age of the solar system.  
We use the latest OPAL opacities \citep{ig96}
for the interior of the star down to temperatures of
$\log T({\rm K}) = 4$. For lower temperatures, we use
the molecular opacities of \citet{af94}.  For regions of
the star with $\log T({\rm K}) \geq 6$, we used the
OPAL equation of state \citep{rsi96}.  For regions where
$\log T(K) \leq 5.5$, we used the equation of state from
\citet{saumon95}, which calculates particle densities
for hydrogen and helium, including partial dissociation
and ionization by both pressure and temperature. In
the transition region between these two temperatures,
both formulations are weighted with a ramp function and
averaged.  The equation of state includes both radiation
pressure and electron degeneracy pressure. For the surface
boundary condition, we experimented with several stellar
atmosphere models as described below.  For our base
case, we adopted $Y = 0.273$ for the Hyades and ignored
diffusion (details
to follow), and used the standard
B\"ohm-Vitense mixing length theory \citep{bv58,cg68}
with $\alpha = 1.72$, to match the solar radius
($R_\odot = 6.9598 \times 10^{8}$ m) and luminosity
($L_\odot = 3.8515 \times 10^{26}$ W) at the
present age of the Sun (4.57 Gyr). 

The evolutionary tracks were generated
for a Hyades metallicity of [Fe/H] $= +0.13$ and
scaled solar abundances \citep{bf90,psc03}.  The tracks were
interpolated to form an isochrone for an adopted
age of 550 Myr, consistent with ages derived from
models excluding convective overshoot \citep{per98}.
Models with overshoot have ages $\approx 625$ Myr, but since we are
dealing with relatively low mass stars, 
the comparison to vB~22 is completely
insensitive to the choice of cluster age.  
The color-temperature
relation in \citet{lej98} was used to generate preliminary
colors\footnote{$V - I$ is on the Cousins system,
while the $V - K_s$ colors use the (short) $K$ band.} 
from the model parameters $M_{\rm bol}$ and $T_{\rm eff}$.
The resulting isochrone is given as Table 1;  the sensitivity
of the colors to the choice of color-temperature
relation is discussed below. Note that this isochrone is
not empirically calibrated to match the photometry of the 
Hyades main sequence, a necessary
procedure discussed at length in our next paper.

The principal theoretical uncertainties in the models 
are the adopted mixture of heavy elements and
the physics chosen for the solar calibration,
in particular whether microscopic diffusion is included
or not.  In the Hyades, the average iron abundance 
[Fe/H] is now determined to high precision ($\pm 0.01$ dex), 
and the relative abundances are
near solar for most of the elements that contribute 
significantly to the internal opacities \citep{psc03}.
The choice of model atmospheres and low-temperature opacities only
affects the position in the HR diagram for effective
temperatures below 3500 K; similar comments apply to the
equation of state.  The superb agreement between theory
and data for helioseismology provides some real confidence
in the accuracy of the ingredients of the models for stars
similar to the Sun, such as vB~22. 

Although the effects of microscopic diffusion do not
matter for the Hyades themselves (the cluster is only 12\%
the age of the Sun), they make a difference for the choice
of a calibrating solar model.  
The net effect of diffusion is a gradual decrease in the
helium and heavy element content in the outer layers with
a smaller fractional rise in the central values of these
quantities.  Compared to models with a uniform composition
profile, solar models including diffusion have a higher
overall helium abundance and require a higher value of
$\alpha$ to match the solar radius and luminosity.

The proper thing to do, therefore, would be to calibrate
the Hyades models using diffusion models for the Sun.  For 
our base case (the isochrone in Table 1),
however, we ignored this and instead used models that
do not follow diffusion and which were calibrated using
models of the homogeneous Sun even though these are
incompatible with seismology.  The principal reason for this
is that the effect of rotationally induced mixing, which
diminishes the effects of diffusion, is difficult to model
for stars much hotter than the Sun where convection zones
are very thin.  

For the base case, which we will call the ``no diffusion''
models, we chose the helium abundance as illustrated in Figure 1.
We assumed that the helium abundance is a function of the 
heavy element content $Z$ is given by $Y = Y_p
+ (\Delta Y/ \Delta Z) Z$, where $Y_p$ is the primordial
helium abundance.  We took $Y_p = 0.245 \pm 0.002$, an
intermediate value between estimates from cosmic nucleosynthesis
and measures in metal-poor \ion{H}{2} regions
\citep[see][and references therein]{bono02,ti02}.  The
solar helium abundance in models lacking diffusion
is $Y_\odot = 0.266 \pm 0.001$;  these require a 
mixing length set by $\alpha = 1.74$.  Adopting these
values of $Y_p$, $Y_\odot$, and $\alpha$ yields
$Y = 0.273$ for the Hyades using the Sun's surface
metal abundance of $Z_\odot = 0.0176$. 

We also computed models using an alternative set of parameters
calibrated on solar models that include diffusion \citep{bpb01}.
Models compatible with helioseismology that include both 
rotational mixing and diffusion have surface abundances of
$Y_{\odot,{\rm surf}} = 0.249 \pm 0.003$ and 
$Z_{\odot,{\rm surf}} = 0.0176$, which imply an initial
composition of $Y_\odot = 0.274$ and $Z_\odot = 0.019 \pm 0.001$ and
$\alpha = 1.85$.  Using these values to extrapolate a model for
the Hyades at $Z({\rm Hyades}) / Z_\odot = 1.35$, we would
derive $Y = 0.280$ for the Hyades.  These will be called the
``diffusion'' models;  these also follow the effects of
microscopic diffusion, even though it does not produce
significant effects at the age of the Hyades.

\section{Photometry, masses, and radii for vB~22}

Table 2 summarizes three high-precision distance estimates
to vB~22, from an orbital parallax \citep{ps88} 
solution\footnote{This value was apparently 
misquoted by \citet{lebreton01} 
in their Table 1, but does not affect their analysis.},
the Hipparcos trigonometric parallax \citep{per98}, and
the kinematic parallax \citep{debruijne01}.  These are all
in excellent agreement. 

The basic photometric data for the vB~22 system
are summarized in Table 3.  The first two rows of 
that table show the \citet{sm87} photometry for 
vB~22 (i.e., both stars together) in $B$, $V$,
and (Cousins) $I_C$, along with 
the derived luminosity ratio in each filter, 
where the errors are taken from
that paper.  Following this are the
apparent magnitudes in each band for the individual
components derived from the luminosity ratios.   Note that
the photometric errors are significant for the secondary and will
be accounted for in the discussion below.  Finally, we derive the
difference in absolute magnitude between the primary and secondary
in each filter, and also include the
absolute magnitudes that would be obtained from the 
kinematic parallax \citep{debruijne01} of vB~22,
again with errors in the individual components.

In Table 4, we compare the model radii and temperatures
at the masses derived for each
component of vB~22 by \citet{tr02} to their
solution (top ten rows
of the table); we also compare the models to the earlier estimates from
\citet{ps88} (last six rows).  The latter values were the ones
used by \citet{lebreton01}.  In the comparison to
the \citet{tr02} solution, we tabulate the properties of the model
both with and without diffusion, while we only
show the no-diffusion case in comparison to \cite{ps88}. The
quantities derived from the model are on the scale where
the Sun has $M_{\rm bol}$ is 4.746 and the radii and effective
temperatures are obtained from the solar-calibrated
helium and mixing length. 

The first thing to note is that in all cases the model
radii at the observed masses are considerably 
different from the radii derived from the analysis 
of the eclipses by $2.5 - 4.5$ times the formal errors
derived from propagating the mass error.
This indicates that there is an inconsistency\footnote{Since
$L \propto R^2T^4_{\rm eff}$, one could match the
models to the data using two of the quantities
$(L, R, T_{\rm eff})$ but not all three simultaneously. Another 
indication that something is the matter is that
both stars formally have the same gravity in the \cite{tr02}
analysis, which contradicts the strong prediction from
theory that the mean density increases with decreasing mass
for main-sequence stars.  The models predict
$\log g({\rm vB~22A}) - \log g({\rm vB~22B}) = -0.11$.} in the
observational determination of mass, radius, and
luminosity compared to the model.  Since luminosity
is determined by a combination of temperature and radius,
we need to examine these separately to pinpoint the
source of this inconsistency.

To quantify the size of the mismatch in temperature,
we also show in Table 4 the model temperatures at the 
observed masses of vB~22A and vB~22B using the 
mass/radius relation in the models.  We also compare
this to the case in which we take the observed radius
as correct, which would imply a larger temperature for
vB~22A where the observed radius is smaller than in
the model, and a smaller temperature for vB~22B.  Here,
the required changes in temperature are considerable,
amounting to $150 - 190$ K.  

The effect of including diffusion
is shown in the top part of that Table for the
\citet{tr02} solution. 
For the same starting helium abundance the effects of the 
precise value of the mixing length are very 
small for luminosity and 
modest even for the effective temperature: 
the diffusion models would be roughly 35 K 
hotter than the no-diffusion models for the primary 
and 13 K hotter for the secondary.  Thus the inconsistency
with the models is not caused by the treatment of diffusion
in the solar calibration.

Because the Hyades has been well studied spectroscopically,
there exist independent estimates of the luminosity/temperature
relation.  In Figure 2, we show spectroscopically derived
temperatures for a subset of stars in the recent study
of Hyades abundances by \citet{psc03}.  Stars with $T < 6000$ K
and with good $BVI_CK_s$ photometry are shown as open
points with error bars.  The values of $M_V$ are derived 
from individual kinematic parallaxes \citep{debruijne01};
errors in this quantity are dominated by distance errors
rather than photometric errors.  PSC to not list individual
temperature errors, so we took $\pm 50$ K as a representative
value, derived from PSC's comparison of their temperatures
to those in previous studies.  The
solid line on that figure is the Hyades isochrone derived in
this paper.  The filled circles show the temperature of
the isochrone at the \cite{tr02} masses.  The filled
triangles display the temperatures that would be found for
the components of vB~22 if the measured radii were correct.
The agreement between the isochrone and the spectroscopic
temperatures is excellent, which shows that the luminosity/radius
relationship in the models is nearly correct, at least 
under the assumption that all Hyades stars have identical
metallicity.  If on the other hand the
luminosity/temperature relation in the models 
were adjusted to match the radii in the \citet{tr02}
solution (vB~22A hotter by 200 K, vB~22B cooler), then 
there would be difference of about 0.15 dex
between the hottest and coolest stars in the PSC sample
(the hotter stars would be come out more metal rich).

In Figure 3, we compare the isochrones to the \citet{tr02}
solution in $M_V$, $B - V$ and $V - I_C$ as a function of mass.
The isochrone is slightly brighter than the data, which indicates
that the distance to vB~22 is underestimated or, as we will
discuss in $\S$ 4, that the helium abundance we adopted
for the Hyades is too high.  The color-temperature relation
in the isochrone differs from the inferred colors of the
binary components, but in this paper we are mainly concerned
with comparing the model luminosities to the data.  

Table 5 summarizes the errors in absolute magnitude,
radius, or effective
temperature that are contributed by different effects.  
To compute the result of the uncertainty in mass,
we assume the \citet{tr02} error of 0.0062
$M_\odot$ for each star.
The errors from [Fe/H] assume $\sigma{\rm [Fe/H]} = 0.05$ dex
per star,
while those listed for the bolometric correction were
computed by taking the largest difference between the
inferred fluxes for that filter between three different color
calibrations (below) and dividing by 2.  We have also computed
errors that would result if the metal abundance were known
with vanishingly small errors;  these are shown in the
rows labeled ``no Z.''  Because the sign of changes
in metallicity is the same for both components the errors
in their relative fluxes are smaller; we give these values
in the row labeled ``B-A.''

We compare the absolute magnitudes of the two
components of vB~22 to the models in Table
6, where we employ models lacking diffusion
but with two alternative color
calibrations, that of \citet{ah95} and \citet{aam96},
in addition to the one employed as our base
case \citep{lej98}.  The comparison is done at
fixed mass.  In general, the different color
calibrations only change the luminosity by a
few hundredths of a magnitude.  We take the scatter
in the luminosities indicating the size of
errors in the bolometric corrections;  these were
shown in Table 5.  We show the effects of including
diffusion or of forcing the temperature scale to
match the observed stellar radii in Table 7. These
models are for the \citet{lej98} color calibration
only.

The agreement between theory and observation is impressive
for the no diffusion models and well within the expected
errors for both the absolute luminosities and the relative
luminosities.  Overall the different color calibrations
agree best for the $V$ and $B$ bands, while there is more
scatter in the predicted $I_C$-band luminosities.  The relative
fluxes are very close to the predicted level for the $V$
and $B$ bands and are mildly inconsistent with the $I_C$ band
fluxes, especially for vB~22B where the models are fainter than
the data by 0.1 mag or so. This comparison 
indicates that the problem is most likely
in the $I_C$ band bolometric corrections rather than in the
$V$ band bolometric corrections.

However, the agreement is not preserved if the effective
temperatures are altered to the values inferred from the
eclipse data.  Essentially, choosing a lower effective
temperature for the secondary drives down the $V$ and $B$
band fluxes while slightly increasing the $I$-band flux.
As a result the relative flux differences in the $V$ and
$B$ bands become much larger.  This result is insensitive
to the metallicity of the Hyades because decreases in the
metallicity affect the luminosity and effective temperature
of both components in the same sense, while the model
radii are insensitive to the metallicity.  We therefore
conclude that the relative fluxes in different bands, the
mass-luminosity relationship, and the radii obtained from
the eclipse data are not consistent with one another.
One of the three must be in error.  Because of the
insensitivity of the mass-luminosity relationship to errors in the
input physics we view it as more likely that there is some
unresolved issue in one of the two other ingredients.
As in previous analyses of this system, we note that there 
is no obvious single change in
the input physics that can reconcile the models with
both components simultaneously.  What we have added is evidence that
these radii are also inconsistent with the flux ratios.

\section{The Helium Abundance of the Hyades}

Because the luminosity of stellar models at fixed mass 
is very sensitive to the helium abundance 
($\partial M_{\rm bol}/ \partial Y = -10$),
we can formally derive an initial helium abundance for
the Hyades: using the \citet{debruijne01} distance to vB~22,
we find $Y = 0.271 \pm 0.006$ for the no-diffusion models.
The models including diffusion are brighter than those
without, indicating that the initial choice of $Y = 0.280$
was too high at the assumed distance.  Correcting those
models brings the estimated helium abundance down to
$Y = 0.271$, showing that the derived helium abundance
is nearly independent of the details of the solar calibration.

If we adopt a primordial helium abundance of 
0.245, the Hyades would
give a slope $\Delta Y / \Delta Z =
1.11 \pm 0.25$, smaller than the values of 1.5 and 1.2
obtained from solar models with and 
without diffusion respectively.  If the
more realistic initial solar helium abundance 
(including diffusion) and the Hyades helium 
abundance are taken at face value, they
suggest a scatter in helium at fixed metal 
abundance of order 0.009; however,
this range is only marginally significant.  
If we take this
as a one sigma error range, it would imply only a 
small resulting error in a cluster distance modulus at
fixed [Fe/H] of order 0.027 magnitudes.

The Hyades helium abundance we derive
is equal within the errors to the solar value $Y = 0.273$
in models excluding diffusion. The 
solar models with diffusion would predict a higher
helium abundance of 0.280, which is 
not favored by the data; however, it is only of 
order $2\sigma$ from the measurement.
Isochrones are less sensitive to changes in 
helium than models of a given mass, with a 
change of 0.01 in helium causing a change in 
$M_V$ at fixed $T_{\rm eff}$ of 0.03.  We can include 
this in our next paper in the error budget for the absolute distances.

This value is higher than that obtained by Lebreton et
al.\ (2001), who found $Y = 0.255 \pm 0.009$.  There
are two reasons for this difference.  First, we adopted
the revised masses of \citet{tr02};  these yield
predicted luminosities that are 0.08 to 0.09 mag
fainter than found by \citet{ps88}, resulting in
a higher helium abundance.  Second, our models employ
a number of ingredients not used by Lebreton et al.;
as they note, using the OPAL equation of state, 
Kurucz model atmospheres rather than a gray atmosphere, 
and a higher mixing length
all increase the inferred value of $Y$. 

The mixture of heavy elements can also have an impact
on the properties of the models.  There have been two
relatively recent revisions of the \citet{gn93}
abundance scale for the Sun which was used in
this paper.  \citet{gs98} have reduced
the CNO abundances, while \citet{asplund00} has
proposed a downward revision of 9.2\% in the zero-point
between the meteoritic and photospheric abundance scales.
This would not alter those abundances not tied to the
meteoritic scale (C, N, O, Ne, etc.) but would affect
important interior opacity sources such as Si and Fe.
Helioseismic tests \citep{bpb01}
indicate that the \citet{gs98} mixture marginally degrades the
agreement with the measured solar convection zone depth,
but by a degree that is consistent with other known
theoretical uncertainties.  The solar helium abundance
is insensitive to the CNO abundances.  The \citet{asplund00}
mixture has a small impact on the solar sound speed, but
the lower iron abundance results in a lower solar helium
abundance.  This is in disagreement with the measured
surface abundance of helium at the 2-3 $\sigma $ level;
\citet{bpb01} were not able to rule this
out because the degree of disagreement is sensitive to
systematic errors in the helium abundance determination
arising from the equation of state.

We have verified that adopting the \citet{gs98} mixture instead
of the \citet{gn93} mixture produces only small changes in the
isochrones (or even the more sensitive tests possible
in vB~22).  The \citet{asplund00} mixture would imply a
Hyades helium abundance (including diffusion) that is
comparable to the \citet{gn93} or 
\citet{gs98} helium abundance inferred
in the absence of diffusion.  We have also explored the
effects of including small deviations from the solar
mix in the Hyades as measured by \citet{psc03}, and found
that the impact would be of the same order as the
difference between the choice of the solar element mix.

\section{Summary and Remarks}

In this paper, we have begun a new analysis of open cluster distances
by performing stringent tests the of the luminosity/temperature for new
isochrones with updated physics.  We explored the effect of varying
many details of the models, including the abundance mix of 
helium and metals, and concluded that the models match the
relative temperatures and luminosities of the components of the
binary vB~22 in the Hyades.  Along the way, we derived a helium
abundance that is not much different from that in the Sun, even
though the Hyades is more metal-rich.

The resulting helium abundance is almost insensitive to the details
of the solar calibration used to generate the isochrone.  Models 
using a solar calibration consistent with helioseismology predict
a high value of $Y$ for the Hyades;  the resulting isochrone, however,
is too bright at fixed mass, which leads to the conclusion that the
Hyades helium abundance is not much different from the Sun's
initial value of $Y = 0.272$.  This may imply a scatter in the
helium abundance of about $\delta Y = 0.01$ at fixed metallicity in
the solar neighborhood.  While this would produce a real effect 
on the luminosity of the isochrones at fixed mass, we have argued
that the effect on the isochrones at fixed color would be much
smaller and would not produce big errors in the distance estimates
derived from main-sequence fitting.

The principal result of this paper was our demonstration that there is
an inconsistency between the solar models and the binary data in
mass, luminosity, temperature, and radius.  External checks on the
abundance and temperature scales from recent high-precision abundances
in the Hyades \citet{psc03}, indicates that the problem lies with
the observational determination of the radii.  We conclude from this
that we can adjust the isochrones to match the Hyades photometry by
leaving the stellar luminosity and temperatures as they are, and
compute corrections to the color/temperature relations;  this is the
subject of our next paper.

The comparison between the model colors and absolute magnitude and
the data for vB~22 may be complicated by the presence of spots on
the stellar surfaces.  In the Pleiades, stars with the luminosities
of vB~22B fall up to 0.5 mag below the main sequence in $V$, $B - V$
as defined by the Hyades \citet{vla87,sta03}, the difference is negligible
when $V$ and $V - I$ are used.  \citet{sta03} attribute this to
the presence of hot and cool regions on the surfaces of the rapidly-rotating
Pleiades stars as compared to their slowly-rotating counterparts
in the Hyades.  The orbital period of vB~22 is 5.6 days \citep{sm87}.
If both members of vB~22 were corotating, this would correspond to equatorial 
rotation speeds of only 8 km s$^{-1}$ for vB~22A and 6 km s$^{-1}$ for vB~22B
(the binaries are well detached).
These values are not much different from the average 
$v \sin i$ for Hyades stars of similar colors in \citet{psc03},
and considerably slower than the projected rotation
speed ($v \sin i$) for the Pleiades stars with the most
anomalous $B - V$ color.  The vB~22 system, however, has exhibited
flares and color variations outside eclipses, so active regions may indeed be
important at some level \citep{sm87}.  

\acknowledgements

The work reported here was supported in part by the National Science
Foundation, under grants AST-9731621 and AST-0206008 to the Ohio State
University Research Foundation.  We wish to thank the anonymous referee
for many helpful comments.

\clearpage
\begin{deluxetable}{lccccc}
\tablewidth{0pt}
\tablecaption{
Theoretical Isochrone for the Hyades
($Y = 0.273$, [Fe/H] $= +0.13$)}
\tablehead{
  \colhead{$M_V$} &
  \colhead{$T_{\rm eff}$} &
  \colhead{$M / M_\odot$} &
  \colhead{$B - V$} &
  \colhead{$V - I_C$} &
  \colhead{$V - K_s$}
}
\startdata
1.07	&	8279	&	2.252	&	0.112	&	0.110	&	0.316	\\
1.20	&	8270	&	2.192	&	0.116	&	0.112	&	0.320	\\
1.35	&	8236	&	2.127	&	0.126	&	0.122	&	0.333	\\
1.50	&	8158	&	2.062	&	0.143	&	0.141	&	0.360	\\
1.65	&	8052	&	1.997	&	0.166	&	0.169	&	0.396	\\
1.80	&	7942	&	1.934	&	0.191	&	0.200	&	0.436	\\
1.95	&	7830	&	1.873	&	0.215	&	0.229	&	0.477	\\
2.10	&	7706	&	1.815	&	0.241	&	0.254	&	0.520	\\
2.25	&	7568	&	1.757	&	0.269	&	0.271	&	0.565	\\
2.40	&	7423	&	1.702	&	0.295	&	0.286	&	0.611	\\
2.55	&	7280	&	1.651	&	0.318	&	0.309	&	0.656	\\
2.70	&	7139	&	1.603	&	0.340	&	0.339	&	0.703	\\
2.85	&	7001	&	1.556	&	0.362	&	0.374	&	0.752	\\
3.00	&	6871	&	1.511	&	0.385	&	0.408	&	0.807	\\
3.15	&	6754	&	1.468	&	0.409	&	0.437	&	0.871	\\
3.30	&	6650	&	1.426	&	0.433	&	0.462	&	0.938	\\
3.45	&	6556	&	1.387	&	0.455	&	0.485	&	1.005	\\
3.60	&	6466	&	1.350	&	0.476	&	0.507	&	1.070	\\
3.75	&	6382	&	1.315	&	0.496	&	0.529	&	1.133	\\
3.90	&	6300	&	1.281	&	0.515	&	0.550	&	1.193	\\
4.05	&	6220	&	1.248	&	0.533	&	0.571	&	1.252	\\
4.20	&	6141	&	1.217	&	0.552	&	0.593	&	1.312	\\
4.35	&	6063	&	1.187	&	0.571	&	0.614	&	1.370	\\
4.50	&	5982	&	1.157	&	0.590	&	0.637	&	1.422	\\
4.65	&	5899	&	1.128	&	0.612	&	0.662	&	1.465	\\
4.80	&	5815	&	1.099	&	0.635	&	0.687	&	1.509	\\
4.95	&	5732	&	1.071	&	0.659	&	0.713	&	1.560	\\
5.10	&	5648	&	1.045	&	0.686	&	0.740	&	1.623	\\
5.25	&	5561	&	1.020	&	0.715	&	0.768	&	1.696	\\
5.40	&	5471	&	0.995	&	0.746	&	0.794	&	1.775	\\
5.55	&	5377	&	0.972	&	0.780	&	0.819	&	1.855	\\
5.70	&	5279	&	0.949	&	0.816	&	0.842	&	1.933	\\
5.85	&	5179	&	0.927	&	0.855	&	0.866	&	2.007	\\
6.00	&	5080	&	0.906	&	0.895	&	0.890	&	2.081	\\
6.15	&	4986	&	0.886	&	0.936	&	0.916	&	2.161	\\
6.30	&	4898	&	0.867	&	0.977	&	0.946	&	2.248	\\
6.45	&	4816	&	0.850	&	1.016	&	0.982	&	2.341	\\
6.60	&	4740	&	0.833	&	1.055	&	1.025	&	2.436	\\
6.75	&	4667	&	0.818	&	1.093	&	1.074	&	2.530	\\
6.90	&	4597	&	0.803	&	1.130	&	1.127	&	2.619	\\
7.05	&	4527	&	0.788	&	1.163	&	1.179	&	2.703	\\
7.20	&	4457	&	0.773	&	1.194	&	1.231	&	2.780	\\
7.35	&	4390	&	0.757	&	1.220	&	1.280	&	2.850	\\
7.50	&	4326	&	0.742	&	1.244	&	1.328	&	2.915	\\
7.65	&	4265	&	0.726	&	1.265	&	1.377	&	2.976	\\
7.80	&	4209	&	0.710	&	1.284	&	1.427	&	3.035	\\
7.95	&	4158	&	0.694	&	1.301	&	1.478	&	3.092	\\
8.10	&	4110	&	0.678	&	1.317	&	1.529	&	3.149	\\
8.25	&	4066	&	0.663	&	1.332	&	1.577	&	3.207	\\
8.40	&	4024	&	0.647	&	1.345	&	1.620	&	3.266	\\
8.55	&	3984	&	0.631	&	1.356	&	1.657	&	3.327	\\
8.70	&	3947	&	0.616	&	1.366	&	1.689	&	3.390	\\
8.85	&	3910	&	0.600	&	1.375	&	1.719	&	3.455	\\
9.00	&	3876	&	0.585	&	1.383	&	1.747	&	3.520	\\
9.15	&	3842	&	0.569	&	1.390	&	1.775	&	3.585	\\
9.30	&	3809	&	0.553	&	1.396	&	1.802	&	3.646	\\
9.45	&	3777	&	0.537	&	1.402	&	1.830	&	3.703	\\
9.60	&	3746	&	0.520	&	1.407	&	1.858	&	3.753	\\
9.75	&	3715	&	0.503	&	1.413	&	1.887	&	3.796	\\
9.90	&	3684	&	0.485	&	1.418	&	1.917	&	3.833	\\
10.05	&	3655	&	0.466	&	1.423	&	1.947	&	3.869	\\
10.20	&	3628	&	0.448	&	1.427	&	1.979	&	3.907	\\
10.35	&	3602	&	0.429	&	1.432	&	2.011	&	3.949	\\
10.50	&	3577	&	0.411	&	1.436	&	2.044	&	3.992	\\
10.65	&	3555	&	0.393	&	1.440	&	2.076	&	4.037	\\
10.80	&	3535	&	0.376	&	1.443	&	2.107	&	4.082	\\
10.95	&	3516	&	0.359	&	1.444	&	2.140	&	4.131	\\
11.10	&	3498	&	0.342	&	1.443	&	2.173	&	4.186	\\
11.25	&	3481	&	0.325	&	1.440	&	2.208	&	4.244	\\
11.40	&	3465	&	0.308	&	1.436	&	2.242	&	4.305	\\
11.55	&	3450	&	0.292	&	1.431	&	2.276	&	4.364	\\
11.70	&	3436	&	0.277	&	1.427	&	2.309	&	4.422	\\
11.85	&	3421	&	0.261	&	1.422	&	2.342	&	4.479	\\
11.96	&	3411	&	0.250	&	1.419	&	2.365	&	4.520	\\
\enddata
\end{deluxetable}


\begin{deluxetable}{lcl}
\tablecaption{Distances to vB~22}
\tablehead{
 \colhead{Reference} &
 \colhead{Distance modulus} &
 \colhead{Method}
}
\startdata
Peterson \& Solensky (1988)                & $3.35  \pm 0.02$  & orbital parallax \\
Perryman et al.\ (1998)                    & $3.348 \pm 0.129$ & trigonometric parallax \\
de Bruijne, Hoogerwerf, \& de Zeeuw (2001) & $3.372 \pm 0.039$ & kinematic parallax \\
\enddata
\end{deluxetable}


\begin{deluxetable}{llccc}
\tablecaption{Photometry of vB~22 and derived quantities\label{vB22data}}
\tablewidth{0pt}
\tablehead{
 \colhead{Quantity} &
 \colhead{Star(s)} &
 \colhead{$V$} &
 \colhead{$B$} &
 \colhead{$I_C$}
}
\startdata 
Photometry               & A+B     & $ 8.319 \pm 0.009$ & $ 9.075 \pm 0.010$ & $7.486 \pm 0.006$\\
$L_{\rm A}/(L_{\rm A}+L_{\rm B})$  
                         & \nodata & $ 0.892 \pm 0.006$ & $ 0.928 \pm 0.004$ & $0.823 \pm 0.008$\\
Photometry               & A       & $ 8.443 \pm 0.011$ & $ 9.156 \pm 0.011$ & $7.697 \pm 0.012$\\
Photometry               & B       & $10.735 \pm 0.061$ & $11.932 \pm 0.061$ & $9.366 \pm 0.050$\\
$M_\lambda(B) - M_\lambda(A)$
                         & \nodata & $ 2.29  \pm 0.06$  & $ 2.78  \pm 0.06$  & $1.67  \pm 0.05$ \\
$M_\lambda$              & A       & $ 5.07  \pm 0.035$ & $ 5.78  \pm 0.035$ & $4.32  \pm 0.035$\\
$M_\lambda$              & B       & $ 7.36  \pm 0.069$ & $ 8.56  \pm 0.069$ & $5.99  \pm 0.060$\\
\enddata
\end{deluxetable}


\begin{deluxetable}{lccl}
\tablewidth{0pt}
\tablecaption{Physical data for vB~22 and models}
\tablehead{
    \colhead{Quantity} &
    \colhead{vb~22A} &
    \colhead{vB~22B} &
    \colhead{Comment}
}
\startdata
$M / M_\odot$ (observed) & $1.0591 \pm 0.0062$& $0.7605 \pm 0.0062$ & TR02 solution \\
$R / R_\odot$ (observed) & $0.900  \pm 0.016$ & $0.768 \pm 0.010$   & TR02 solution \\
$R / R_\odot$ (model)    & $0.962  \pm 0.007$ & $0.714 \pm 0.007$   & no diffusion \\
$R / R_\odot$ (model)    & $0.955  \pm 0.007$ & $0.712 \pm 0.007$   & with diffusion \\
$T_{\rm eff}$ (model)    & $5680 \pm 60$      & $4400 \pm 60$       & no diffusion, at listed mass \\
$T_{\rm eff}$ (model)    & $5870 \pm 50$      & $4240 \pm 30$       & no diffusion, if observed radius is correct \\
$T_{\rm eff}$ (model)    & $5780 \pm 60$      & $4460 \pm 60$       & with diffusion, at listed mass \\
$T_{\rm eff}$ (model)    & $5950 \pm 60$      & $4290 \pm 30$       & with diffusion, if TR02 radius is correct \\
$M_{\rm bol}$ (model)    & 4.90               & 6.66                & no diffusion \\
$M_{\rm bol}$ (model)    & 4.83               & 6.59                & with diffusion \\
 \\
$M / M_\odot$ (observed) & $1.072 \pm 0.062$  & $0.769 \pm 0.005$   & PS88 solution \\
$R / R_\odot$ (observed) & $0.905 \pm 0.029$  & $0.773 \pm 0.015$   & PS88 solution \\
$R / R_\odot$ (model)    & $0.974  \pm 0.007$ & $0.714 \pm 0.007$   &  \\
$M_{\rm bol}$ (model)    & 4.84               & 6.61                &  \\
$T_{\rm eff}$ (model)    & $5930 \pm 50$      & $4290 \pm 40$       & if PS88 radius is correct \\
$T_{\rm eff}$ (model)    & $5720 \pm 60$      & $4400 \pm 60$       & at listed mass \\
\enddata
\tablecomments{PS88 = \citet{ps88}.  TR02 = \citet{tr02}.  All the comparisons with PS88
assume no diffusion.}
\end{deluxetable}


\begin{deluxetable}{lcccccccccc}
\tablecaption{Error analysis}
\tablehead{
   \colhead{} &
   \multicolumn{5}{c}{Star A} &
   \multicolumn{5}{c}{Star B} \\
   \colhead{Theoretical errors} &
   \colhead{$V$} &
   \colhead{$B$} &
   \colhead{$I_C$} &
   \colhead{$R/R_\odot$} &
   \colhead{$T_{\rm eff}$} &
   \colhead{$V$} &
   \colhead{$B$} &
   \colhead{$I_C$} &
   \colhead{$R/R_\odot$} &
   \colhead{$T_{\rm eff}$}
}
\startdata
$\sigma$(mass)    & 0.03 & 0.03 & 0.04 & 0.006 & 18 & 0.06 & 0.07 & 0.04 & 0.006 & 27 \\
$\sigma$([Fe/H])  & 0.06 & 0.08 & 0.04 & 0.003 & 58 & 0.09 & 0.12 & 0.06 & 0.003 & 56 \\
$\sigma$(B.C.)    & 0.02 & 0.03 & 0.03 & \nodata & \nodata & 0.03 & 0.01 & 0.04 & \nodata & \nodata \\
$\sigma$(total)   & 0.07 & 0.09 & 0.06 & 0.007 & 61 & 0.11 & 0.14 & 0.08 & 0.007 & 62 \\
$\sigma$(total) -- no Z & 0.04 & 0.04 & 0.05 & 0.006 & 18 & 0.07 & 0.07 & 0.06 & 0.007 & 62 \\
$\sigma$(total) (B -- A) & 0.08 & 0.09 & 0.08 & \nodata & \nodata & \nodata & \nodata & \nodata & \nodata & \nodata\\
\enddata
\end{deluxetable}


\begin{deluxetable}{lllllllll}
\tablewidth{0pt}
\tablecaption{Effect of color calibration on no-diffusion model}
\tablehead{
  \colhead {} & 
  \colhead {} &
  \multicolumn{3}{c}{Color calibration} &
  \colhead {} & 
  \multicolumn{3}{c}{Difference (model -- data)} \\
  \colhead{Star} &
  \colhead{Quantity} &
  \colhead{A\&H\tablenotemark{a}} &
  \colhead{AAM\tablenotemark{b}} &
  \colhead{LCB\tablenotemark{c}} &
  \colhead{Data} &
  \colhead{A\&H} &
  \colhead{AAM} &
  \colhead{LCB}
}
\startdata
vB~22A & $M_V$ & 5.02 & 5.05 & 5.03 & $5.07 \pm 0.04$ & $-0.05$ & $-0.02$ & $-0.04$ \\
       & $M_B$ & 5.73 & 5.71 & 5.71 & $5.78 \pm 0.04$ & $-0.05$ & $-0.07$ & $-0.02$ \\
       & $M_I$ & 4.29 & 4.36 & 4.30 & $4.32 \pm 0.04$ & $-0.03$ & $+0.04$ & $-0.02$ \\
       & $B-V$ & 0.71 & 0.66 & 0.68 & $0.71 \pm 0.06$ & & & \\
       & $V-I$ & 0.73 & 0.69 & 0.73 & $0.75 \pm 0.06$ & & & \\
vB~22B & $M_V$ & 7.37 & 7.38 & 7.34 & $7.36 \pm 0.07$ & $+0.01$ & $+0.02$ & $-0.02$ \\
       & $M_B$ & 8.53 & 8.52 & 8.53 & $8.56 \pm 0.07$ & $-0.03$ & $-0.04$ & $-0.03$ \\
       & $M_I$ & 6.09 & 6.15 & 6.10 & $5.99 \pm 0.06$ & $+0.10$ & $+0.16$ & $+0.11$ \\
       & $B-V$ & 1.20 & 1.14 & 1.19 & $1.20 \pm 0.07$ & & & \\
       & $V-I$ & 1.37 & 1.23 & 1.24 & $1.37 \pm 0.07$ & & & \\ 
\enddata
\tablenotetext{a}{Allard \& Hauschildt (1995).}
\tablenotetext{b}{Alonso, Arribas, \& Martinez Roger (1996).}
\tablenotetext{c}{Lejeune, Cuisinier, \& Buser (1998).}
\end{deluxetable}


\begin{deluxetable}{llcccc}
\tablewidth{0pt}
\tablecaption{Alternative models}
\tablehead{
  \colhead {} & 
  \colhead {} &
  \multicolumn{2}{c}{Alternative model\tablenotemark{a}} &
  \multicolumn{2}{c}{Difference (model -- data)} \\
  \colhead{Star} &
  \colhead{Quantity} &
  \colhead{diffusion} &
  \colhead{force radius} &
  \colhead{diffusion} &
  \colhead{force radius} 
}
\startdata
vB~22A & $M_V$ & 5.00 & 4.97 & $-0.07$ & $-0.10$ \\
       & $M_B$ & 5.65 & 5.62 & $-0.13$ & $-0.16$ \\
       & $M_I$ & 4.34 & 4.28 & $+0.02$ & $-0.04$ \\
       & $B-V$ & 0.65 & 0.65 & & \\
       & $V-I$ & 0.66 & 0.69 & & \\
vB~22B & $M_V$ & 7.48 & 7.31 & $+0.12$ & $-0.05$ \\
       & $M_B$ & 8.73 & 8.47 & $-0.17$ & $-0.09$ \\
       & $M_I$ & 6.14 & 5.92 & $+0.22$ & $+0.15$ \\
       & $B-V$ & 1.25 & 1.16 & & \\
       & $V-I$ & 1.34 & 1.39 & & \\ 
\enddata
\tablenotetext{c}{Lejeune, Cuisinier, \& Buser (1998) color calibration.}
\end{deluxetable}

\clearpage

\figcaption[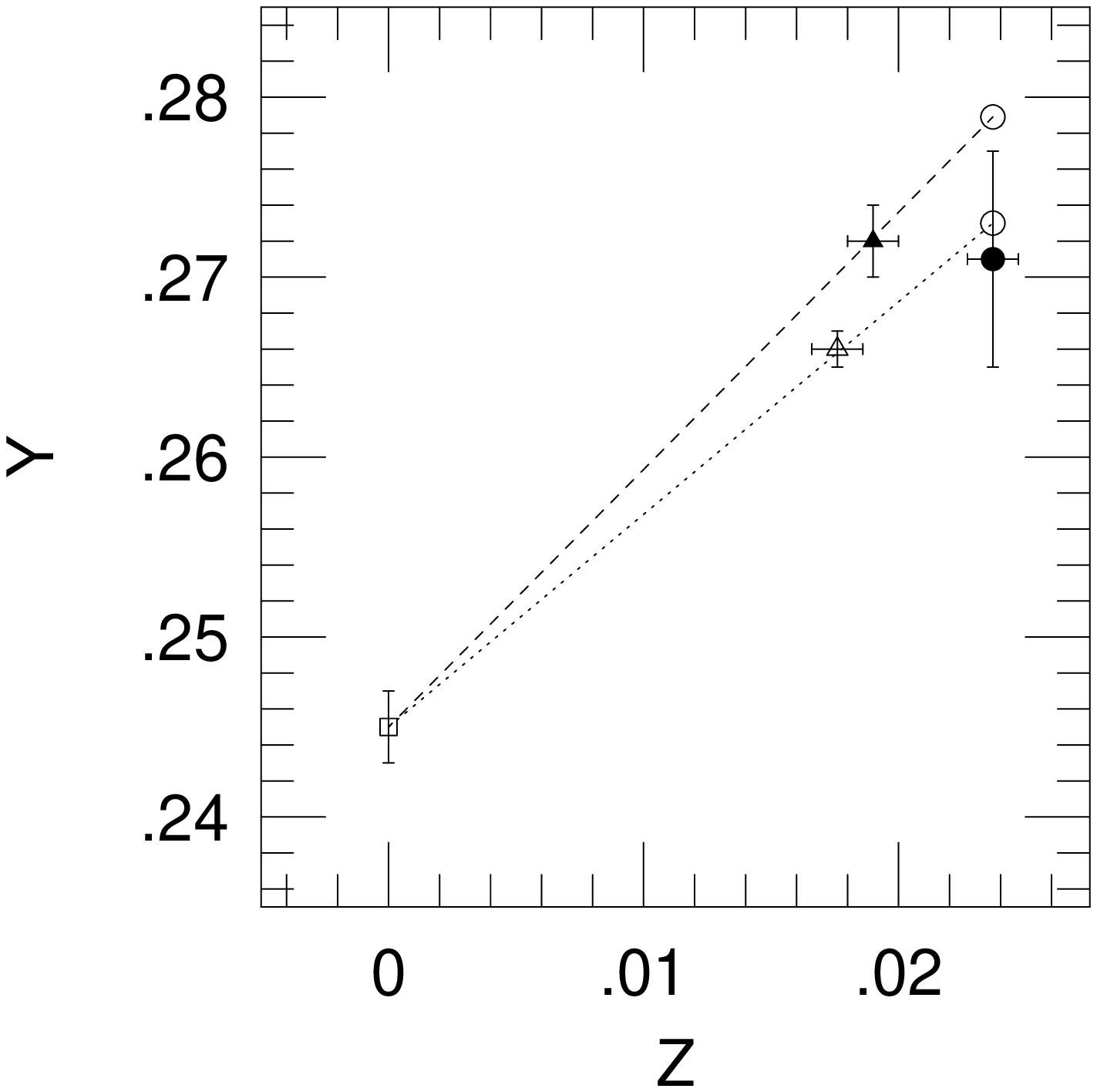]{Derivation of the helium abundance $Y$ for the
Hyades isochrone.  The lower dashed line shows the extrapolation
from the primordial helium abundance through models of the Sun that
lack diffusion to the Hyades metal abundance [Fe/H] $= + 0.13 \pm 0.01$
($Z = 0.0237$);
this is the base case discussed in the paper and shown in Table 1.
The upper line shows the extrapolation through solar models consistent
with helioseismology, which would imply a higher helium abundance in the
Hyades.  Both initial values are shown as open circles.  The filled
point shows the helium abundance for the Hyades found in this
paper.}

\figcaption[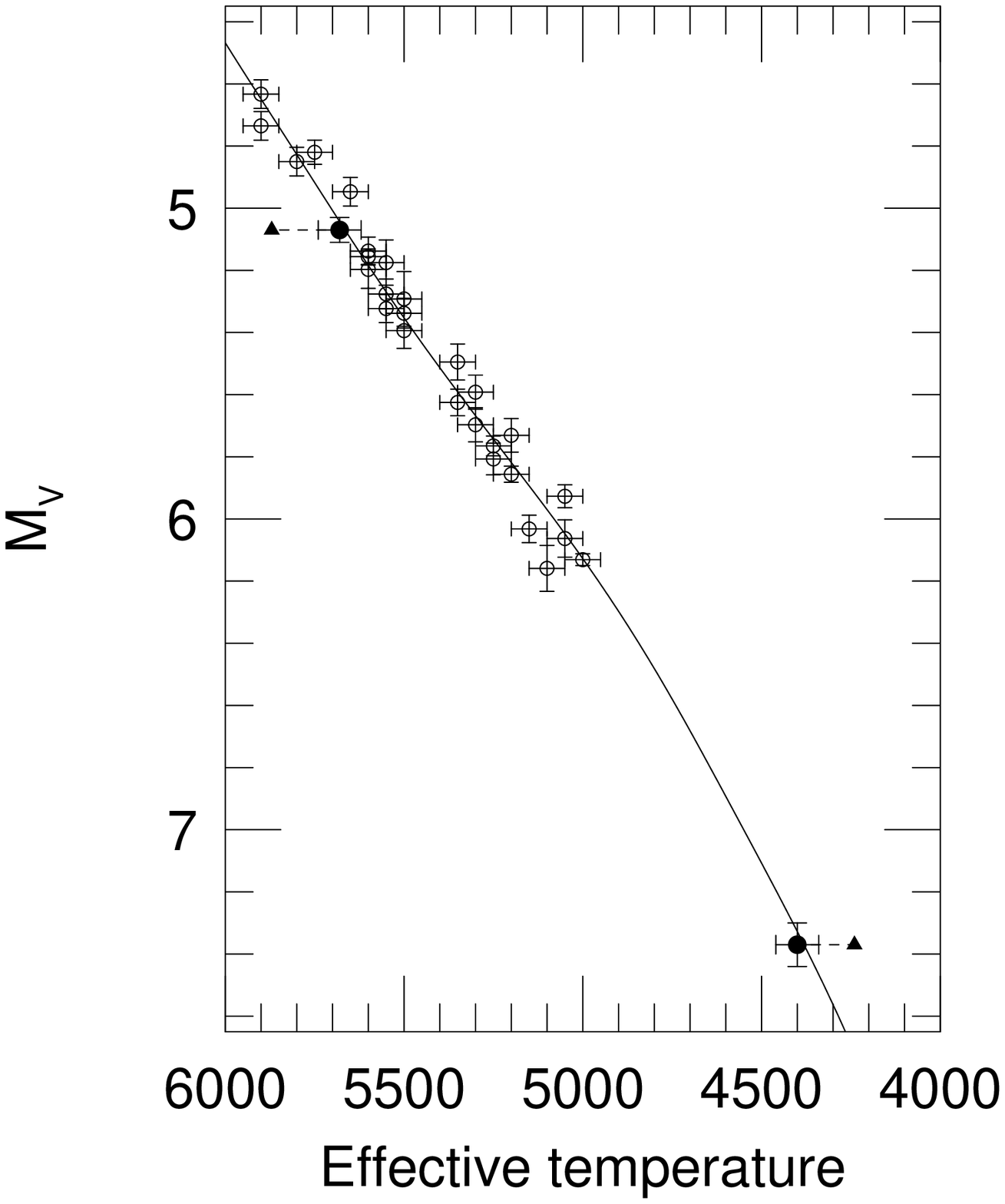]{Comparison of isochrone and spectroscopic
temperature scales for the Hyades.  The solid line is the
theoretical isochrone in Table 1, while the open points with
error bars show spectroscopic temperatures from Paulson, Sneden,
\& Cochran (2003) along with absolute visual magnitude computed
from Hyades kinematic parallaxes (de Bruijne, Hoogerwerf, \& de Zeeuw
(2001).  The filled circles are for vB~22A and vB~22B at the
isochrone temperature and
measured masses for each component, while the triangles indicate the
temperatures that would be derived by forcing the models to
have the radius indicated by the Torres \& Ribas (2002) solution. }

\figcaption[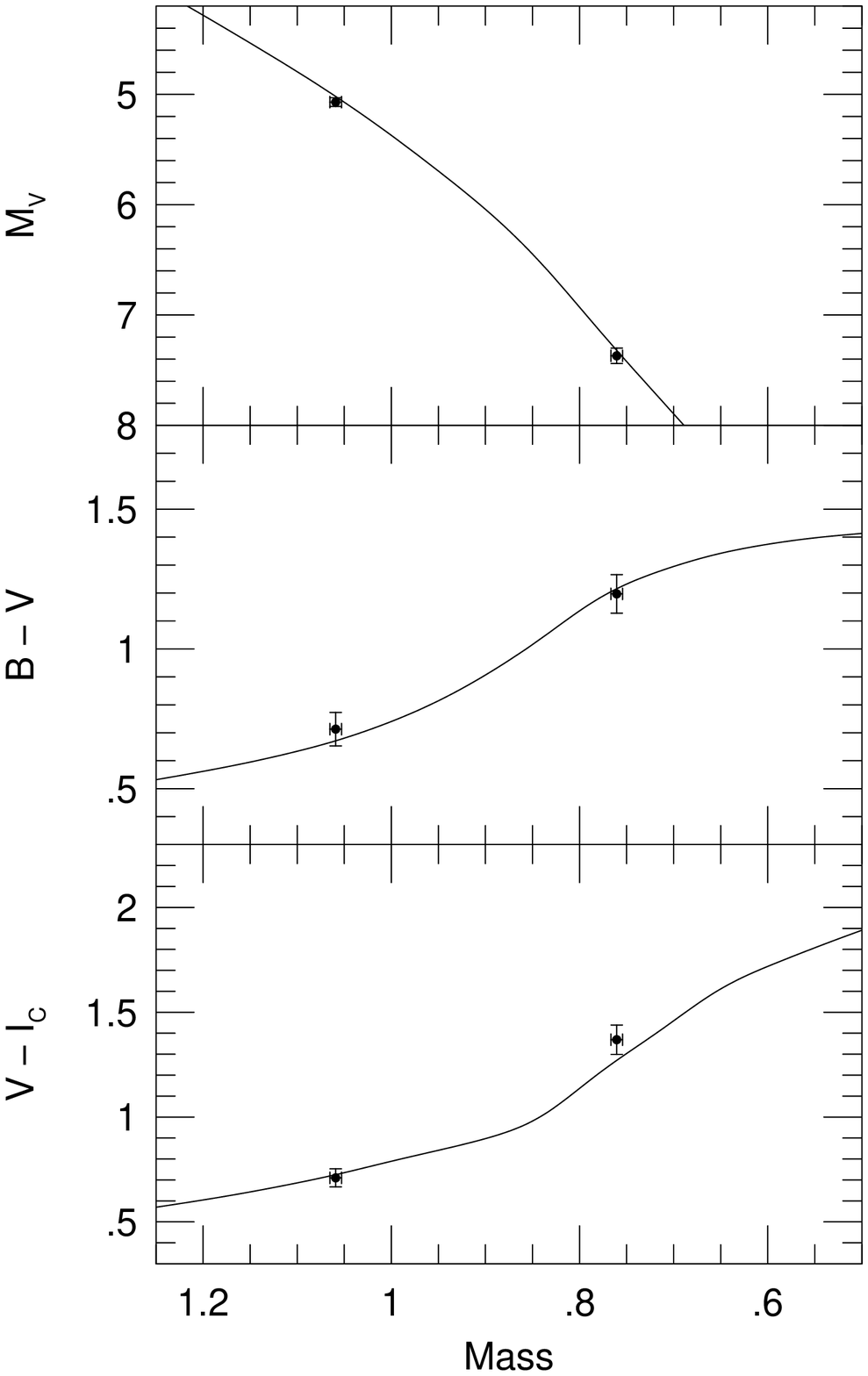]{Comparison of the isochrone to data for vB~22.
The isochrone is shown as a solid line, and the values for vB~22A and
vB~22B are shown as points with error bars.}


\begin{thebibliography}


\bibitem[Alexander \& Ferguson(1994)]{af94} Alexander,
D. R., \& Ferguson, J. W. 1994, \apj, 437, 879

\bibitem[Allard \& Hauschildt(1995)]{ah95} Allard, F., \&
Hauschildt, P. H. 1995, \apj, 445, 433

\bibitem[Alonso, Arribas, \& Mart\'{\i}nez-Roger(1996)]{aam96}
Alonso, A., Arribas, S., \& Mart\'{\i}nez-Roger, C. 1996,
\aap, 313, 873

\bibitem[Asplund(2000)]{asplund00} Asplund, M. 2000, \aap,
359, 755

\bibitem[Bahcall, Pinsonneault, \& Basu(2001)]{bpb01}
Bahcall, J. N., Pinsonneault, M. H., \& Basu, S. 2001,
\apj, 555, 990

\bibitem[Boesgaard \& Friel(1990)]{bf90} Boesgaard, A. M., 
\& Friel, E. D. 1990, \apj, 351, 467

\bibitem[B\"ohm-Vitense(1958)]{bv58} B\"ohm-Vitense,
E. 1958, Z.\ Astrophys., 46, 108

\bibitem[Bono et al.(2002)]{bono02} Bono, G., Balbi, A., Cassisi, S., 
Vittorio, N., \& Buonanno, R. 2002, \apj, 568, 463

\bibitem[de Bruijne, Hoogerwerf, \& de Zeeuw(2001)]{debruijne01} 
de Bruijne, J. H. J., Hoogerwerf, R., \& de Zeeuw,
P. T. 2001, \aap, 367, 111 

\bibitem[Castellani, Degl'Innocenti, \& Prada Moroni(2001)]{cas01}
Castellani, V., Degl'Innocenti, S., \& Prada Moroni, P. G.
2001, \mnras, 320, 66

\bibitem[Cox \& Guili(1968)]{cg68} Cox, J. P., \& Guili,
R. T.  1968, Principles of Stellar Structure (New York:
Gordon and Breach)

\bibitem[Grevesse \& Noels(1993)]{gn93} Grevesse, N., \& Noels,
A. 1993 in Origin and Evolution of the
Elements, ed. M. Prantzos, E. Vangioni-Flam, \& M. Cass\'e
(Cambridge:  Cambridge Univ. Press), 15

\bibitem[Grevesse \& Sauval(1998)]{gs98} Grevesse, N., \& Sauval,
A. J. 1993, Space Sci.\ Rev., 85, 161

\bibitem[Gruzinov \& Bahcall(1998)]{gb98} Gruzinov, A.,
\& Bahcall, J. 1998, \apj, 504, 996

\bibitem[Guenther et al.(1992)]{guenther92} Guenther,
D. B., Demarque, P., Kim, Y.-C., \& Pinsonneault,
M. H. 1992, \apj, 387, 372

\bibitem[Iglesias \& Rogers(1996)]{ig96} Iglesias, C. A.,
\& Rogers, F. J. 1996, \apj, 464, 943

\bibitem[Jimenez et al.(2003)]{jimenez03} Jimenez, R., Flynn, C.,
MacDonald, J., \& Gibson, B. K. 2003, preprint (astro-ph/0303179)

\bibitem[Lastennet et al.(1999)]{lastennet99} Lastennet, E., 
Valls-Gabaud, D., Lejeune, T., \& Oblak, E. 1999, \aap, 349, 485

\bibitem[Lebreton, Fernandes, \& Lejeune(2001)]{lebreton01} Lebreton, 
Y., Fernandes, J., \& Lejeune, T. 2001, \aap, 374, 540

\bibitem[van Leeuwen(1999)]{vl99} van Leeuwen, F. 1999, \aap, 341, 71

\bibitem[van Leeuwen, Alphenaar, \& Meys(1987)]{vla87}
van Leeuwen, F., Alphenaar, P., \& Meys, J. J. M. 1987, \aap, 67, 483

\bibitem[Lejeune, Cuisinier, \& Buser(1998)]{lej98} Lejeune, T.,
Cuisinier, F., \& Buser, R. 1998, \aap, 313, 873

\bibitem[Makarov(2002)]{makarov02} Makarov V. V., 2002, \aj, 124, 3299

\bibitem[McClure(1982)]{mcclure82} McClure, R. D. 1982, \apj, 254, 606

\bibitem[Paulson, Sneden, \& Cochran(2003)]{psc03} Paulson, 
D. B., Sneden, C., \& Cochran, W. D. 2003, \aj, 125, 3185

\bibitem[Percival, Salaris, \& Kilkenny(2003)]{per03} Percival, 
S. M., Salaris, M., \& Kilkenny, D. 2003, \aap, 400, 541

\bibitem[Perryman et al.(1998)]{per98} Perryman, M. A. C., et al.
1998, \aap, 331, 81

\bibitem[Peterson \& Solensky(1988)]{ps88}
Peterson, D. M., \& Solensky, R. 1988, \apj, 333, 256

\bibitem[Radick et al.(1987)]{radick87} Radick, R. R., 
Thompson, D. T., Lockwood, G. W., Duncan, D. K., \& 
Baggett, W. E. 1987, \apj, 321, 459

\bibitem[Robichon et al.(1999)]{rob99} Robichon, N., Arenou, F.,
Mermilliod, J.-C., \& Turon, C. 1999, \aap, 345, 471

\bibitem[Rogers, Swenson, \& Iglesias(1996)]{rsi96}
Rogers, F. J., Swenson, F. J., \& Iglesias, C. A. 1996,
\apj, 456, 902

\bibitem[Saumon, Chabrier, \& Van Horn(1995)]{saumon95}
Saumon, D., Chabrier, G., \& Van Horn, H. M. 1995, \apjs,
99, 713

\bibitem[Schiller \& Malone(1987)]{sm87} Schiller,
S. J., \& Malone, E. F. 1987, \aj, 93, 1471

\bibitem[Sills, Pinsonneault, \& Terndrup(2000)]{spt00}
Sills, A., Pinsonneault, M. H., \& Terndrup, D. M. 2000,
\apj, 534, 335

\bibitem[Stauffer et al.(2003)]{sta03} Stauffer, J. R., Jones,
B. F., Backman, D., Hartmann, L. W., Barrado y Nevascu\'es, D.,
Pinsonneault, M. H., Terndrup, D. M., \& Muench, A. 2003,
preprint

\bibitem[Terndrup et al.(2000)]{ter00} Terndrup, D. M., Stauffer,
J. R., Sills, A., Yuan, Y., Jones, B. F., Fischer, D.,  \&
Krishnamurthi, A. 2000, \apj, 119, 1303

\bibitem[Thuan \& Izotov(2002)]{ti02} Thuan, T. X., \& Izotov,
Y. I. 2002, Space Sci. Rev., 100, 263

\bibitem[Torres \& Ribas(2002)]{tr02} Torres, G., \& Ribas, I. 2002, 
\apj, 567, 1140

\end{thebibliography}
\end{document}